\documentclass[10pt, conference, anonymous]{IEEEtran}
\IEEEoverridecommandlockouts
\usepackage{amsmath,amssymb,amsfonts}
\usepackage{graphicx}
\usepackage{textcomp}
\usepackage{xcolor}
\usepackage[hidelinks]{hyperref}
\usepackage{comment}
\usepackage{algorithm}
\usepackage[utf8]{inputenc}
\usepackage{algpseudocode}
\usepackage{amssymb}
\usepackage{booktabs}
\usepackage{multirow}
\usepackage{tcolorbox}
\usepackage{array}
\usepackage{enumitem}
\usepackage[normalem]{ulem}
\useunder{\uline}{\ul}{}
\newboolean{showcomments}
\setboolean{showcomments}{true}
\ifthenelse{\boolean{showcomments}}
{ }
\ifthenelse{\boolean{showcomments}}
{ }
\ifthenelse{\boolean{showcomments}}
{ }
\ifthenelse{\boolean{showcomments}}
{ }

\def\BibTeX{{\rm B\kern-.05em{\sc i\kern-.025em b}\kern-.08em
    T\kern-.1667em\lower.7ex\hbox{E}\kern-.125emX}}

\begin{document}

\title{Can Identifier Splitting Improve Open-Vocabulary Language Model of Code?}

\author{
\IEEEauthorblockN{Jieke Shi, Zhou Yang, Junda He, Bowen Xu\textsuperscript{$\ast$}\thanks{$\ast$ Corresponding author.}, David Lo}
\IEEEauthorblockA{
\textit{School of Computing and Information Systems, Singapore Management University}\\
\{jiekeshi, zyang, jundahe, bowenxu.2017, davidlo\}@smu.edu.sg}
}

\maketitle

\begin{abstract}
Statistical language models on source code have successfully assisted software engineering tasks. However, developers can create or pick arbitrary identifiers when writing source code. Freely chosen identifiers lead to the notorious {\em out-of-vocabulary} (OOV) problem that negatively affects model performance. Recently, Karampatsis et al. showed that using the Byte Pair Encoding (BPE) algorithm to address the OOV problem can improve the language models' predictive performance on source code. However, a drawback of BPE is that it cannot split the identifiers in a way that preserves the meaningful semantics. Prior researchers also show that splitting compound identifiers into sub-words that reflect the semantics can benefit software development tools. These two facts motivate us to explore whether identifier splitting techniques can be utilized to augment the BPE algorithm and boost the performance of open-vocabulary language models considered in Karampatsis et al.'s work. 
	
This paper proposes to split identifiers in both constructing vocabulary and processing model inputs procedures, thus exploiting three different settings of applying identifier splitting to language models for the code completion task. We contrast models' performance under these settings and find that simply inserting identifier splitting into the pipeline hurts the model performance, while a hybrid strategy combining identifier splitting and the BPE algorithm can outperform the original open-vocabulary models on predicting identifiers by $3.68\%$ of recall and $6.32\%$ of Mean Reciprocal Rank. The results also show that the hybrid strategy can improve the entropy of language models by $2.02\%$.
\end{abstract}


\section{Introduction}
\label{sec:intro}

Numerous works have applied statistical language models (LMs) on source code to help tackle important tasks in software engineering, including code completion~\cite{Raychev2014statistical}, program repair~\cite{s2s}, and many others~\cite{survey}. Same as modeling natural language, creating appropriate vocabulary is a crucial prerequisite~\cite{VOLT}. However, when writing source code, software developers can create arbitrary identifiers they like, which probably contain multiple words, e.g. {\tt addItemsToList}. Due to this fundamental fact, models of code often get an extremely sparse vocabulary containing many rare words when processing code corpora. Training models with such sparse (and typically large) vocabulary is ineffective, and obtained models often have poor performance~\cite{VOLT}. In addition, if identifiers are not observed in the vocabulary, the model cannot handle them, which is known as the notorious {\em out-of-vocabulary} (OOV) problem.

Currently, open-vocabulary methods like Byte-Pair Encoding (BPE) algorithm~\cite{sennrich-etal-2016-neural} are widely used in modeling natural languages and achieve promising results in practice. These methods can solve the OOV problem while customizing the size of the vocabulary. Inspired by such success, Karampatsis et al.~\cite{BigCode2020Karampatsis} first applied the BPE algorithm to construct vocabulary from source code and showed that open-vocabulary LMs have outstanding performance on the code completion task. However, BPE selects the most frequent sub-words into the vocabulary, and this frequency-based approach often fails to capture the semantics and intentions of identifier names when choosing sub-words. Although developers create any identifiers at will, they usually follow certain naming conventions that make identifiers meaningful, legible and easy to understand, either in {\em camelCase} or in {\em snake\_case}~\cite{tocamelcase}. For example, the method name {\tt getListener} follows the {\em camelCase} convention, and a programmer can easily infer that this method can be used to {\tt get} a {\tt Listener} object. At the same time, the BPE algorithm will represent it as three sub-words in our preliminary study: {\tt get}, {\tt List} and {\tt ener}, the latter two do not reflect the semantics developers try to convey.

In order to empower the BPE algorithm with the ability to better sense semantics when splitting words, an intuitive preprocessing strategy is to split compound identifiers into several words that can imply certain meanings, which is called {\em identifier splitting} techniques. Prior research works have demonstrated that various information retrieval models for program comprehension tasks are benefit from identifier splitting, e.g., feature-related code localization~\cite{canIS}, code reuse~\cite{codereuse}. However, this empirical conclusion is ambiguous for modern LMs of code as no such work has demonstrated it. As stated above, open-vocabulary LMs assist many software engineering tasks effectively but are weak in capturing the semantics of identifiers when creating vocabulary. Thus, it is imperative to clarify if we can improve the performance at a more considerable margin by combining identifier splitting with the BPE algorithm.

In this paper, we investigate the potential benefits of splitting identifiers in open-vocabulary LMs of code. Specifically, we adopt the same LMs presented by Karampatsis et al.~\cite{BigCode2020Karampatsis}, which are the first to adopt the BPE algorithm in code modeling. To achieve the goal, we propose to apply identifier splitting in two stages of open-vocabulary LMs: vocabulary construction and model input processing. Furthermore, we propose two different preprocessing strategies in these stages to apply identifier splitting techniques: (1) {\it simple strategy}: we split all identifiers in vocabulary construction and apply identifier splitting before in model input processing stages; (2) {\it hybrid strategy}: in the vocabulary construction stage, we first split all identifiers and then merge them with original corpora. In the model input processing stage, we apply identifier splitting only when BPE fails to tokenize them as the original forms. We train LMs under these different settings and evaluate them on the code completion task as~\cite{BigCode2020Karampatsis} to show the effectiveness of identifier splitting in open-vocabulary LMs.

We perform experiments on the C language dataset released by Karampatsis et al.~\cite{BigCode2020Karampatsis}. We evaluate the {\it cross entropy} of LMs, and use {\it Mean Reciprocal Rank} (MRR) to measure the performance of LMs on the code completion task. Furthermore, we also obtain the MRR and {\it recall at rank 10} (R@10) on predicting identifier tokens (excluding keywords, punctuations, etc.). The experimental results show that simply performing identifier splitting into preprocessing procedures does not suffice; it degrades MRR by $0.46\%$ and $5.68\%$ on predicting all tokens and identifiers, respectively. At the same time, the hybrid strategy is more effective for open-vocabulary LMs, outperforming the LMs with the original setting by $6.23\%$ of MRR on predicting identifier tokens. The improvements of $2.02\%$ of LMs entropy and $3.68\%$ in terms of R@10 on predicting identifiers also confirm the hybrid strategy's effectiveness. 
The results highlight that the identifier splitting can be combined with open-vocabulary methods to enhance the performance of language models of source code.

The rest of this paper is organized as follows. Section~\ref{sec:preliminary} briefly describes backgrounds of this paper. In Section~\ref{sec:methodology}, we elaborate our methodology to apply identifier splitting in the code completion pipelines. We describe the experiment settings and present the results of our experiments in Section~\ref{sec:experiment}. Section~\ref{sec:related_work} discusses some related works. Finally, we conclude the paper and present future work in Section~\ref{sec:conclusion}.
\section{Background} 
\label{sec:preliminary}
This section introduces the background of related techniques in this work, including identifier splitting techniques and the Byte-Pair Encoding (BPE) algorithm.

\subsection{Naming Convention and Identifier Splitting}
While identifier names do not affect program functions and human developers can pick or create names at will, some naming conventions are encouraged to follow as they can give meaningful and well-readable names so that other programmers' comprehension of code can be considerably improved~\cite{ClearID}. Two widely adopted naming conventions are {\em camelCase} rule and {\em snake\_case} rule~\cite{tocamelcase}.
However, simply splitting by conventions is not accurate enough when extracting meaningful sub-words from compound identifiers, especially for identifiers that do not strictly follow naming conventions (e.g., the same-case identifier where all characters are in the single case like {\tt httprequest}). Several more precise and innovative identifier splitting approaches are proposed to handle more sophisticated situations beyond conventions, in which Ronin~\cite{Spiral} is the state-of-the-art. Ronin splits identifiers into sub-words based on various heuristic rules and a pre-defined frequent sub-word table. In this paper, we take Ronin as a representative of identifier splitting tools.

\subsection{Byte-Pair Encoding}
In this paper, we use a popular open-vocabulary method called the {\em Byte-Pair Encoding} (BPE) algorithm \cite{sennrich-etal-2016-neural}. BPE algorithm consists of two components: (1) the vocabulary construction stage, which takes text corpora and returns a vocabulary with the predefined size; and (2) the tokenization stage, which segments and tokenizes new corpora with the built vocabulary and returns a sequence of tokens. Initially, corpora are split into a set of tokens while each token only contains one character. The BPE algorithm iteratively merges the most frequent pair of tokens into a new single token until a given maximum number of merge operations is reached. These single tokens are then used to replace the original pair of tokens in the corpora. The resulting vocabulary is an ordered list of sub-word units created from the merge operations. When we run the BPE algorithm on new corpora using the vocabulary, the tokens are merged in the same order as occurred during vocabulary construction.

\section{Methodology} \label{sec:methodology}

% This section illustrates our idea of combining the identifier splitting method and BPE algorithm in the pipeline of LMs of code. We consider three different settings: original, simple splitting, and hybrid strategy as stated in Section~\ref{sec:intro}. The differences between these settings are discussed detailedly in this section.

% Figure \ref{fig:pipeline} illustrates the overview of the code completion task and how the open vocabulary method (i.e., BPE) is used. BPE involves two parts in this pipeline. On the one hand, we use a corpus to create a BPE vocabulary. On the other hand, to handle the OOV problem, BPE is applied to decompose model inputs into lists of {\em subtokens}, which are then fed into the code completion model. Now we discuss how identifier splitting can be used in this pipeline, as well as the metrics to evaluate the impact of adopting identifier splitting. 

\begin{figure*}[t!]
	\centering
	\includegraphics[width=0.78\linewidth]{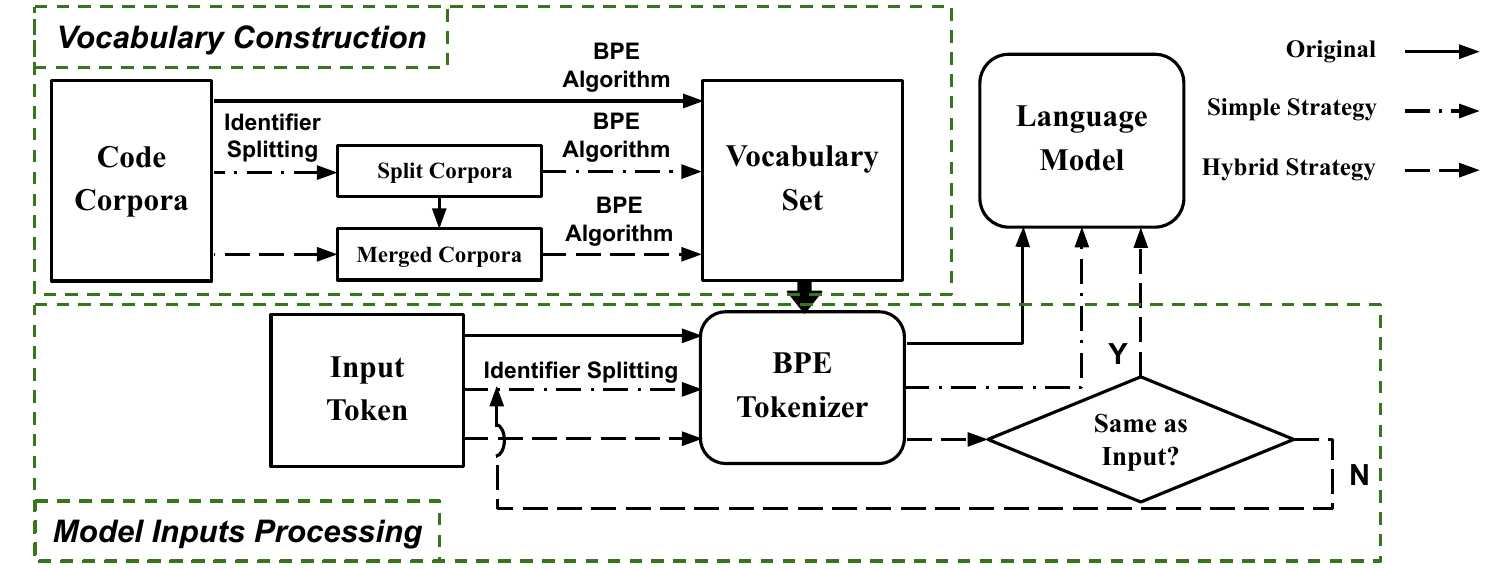}
	\caption{The overview of three strategies about applying the BPE algorithm and identifier splitting in the language models. {\tt Original} refers to the original setting, which only adopts BPE, and no identifier splitting is involved. {\tt Simple Strategy} and {\tt Hybrid Strategy} refer to two different strategies which combine identifier splitting with BPE.}
	\label{fig:pipeline}
\end{figure*}

Figure~\ref{fig:pipeline} presents an overview of how we apply identifier splitting in the open-vocabulary language models. It shows that identifier splitting can be used in both vocabulary construction and model inputs processing procedures before training the language models. We elaborate the motivation and processing details as follows.

\vspace{0.4em}
\noindent{\bf Vocabulary Construction.} 
As shown in Figure~\ref{fig:pipeline}, we apply the BPE algorithm on the input code corpora and output a vocabulary with a predefined size. The vocabulary will be used in the tokenization of model inputs later.

As introduced in Section~\ref{sec:preliminary}, the BPE algorithm builds a vocabulary based on the frequency of pairs appearing in training corpora. Prior works sample some software projects as a corpus to create a vocabulary, and then use the vocabulary to tokenize other mutually exclusive projects~\cite{BigCode2020Karampatsis}. We notice the fact that many complex identifiers are usually project-specific or even file-specific. Researchers have shown that code has a high degree of localness, where identifiers are repeated often within close distance while they are rarely used in other projects~\cite{localness}.
It means that the vocabulary created on one corpus may contain many complex compounds (as they appear very frequently in the corpus). However, those complex compounds are likely to be rare in other corpora. For a vocabulary with limited size, these rare compounds infringe the space of sub-words that could be applied in the BPE procedure of other corpora, reducing the efficiency of the vocabulary. It motivates us to apply identifier splitting techniques to the corpus. 

% Besides, considering the scalability of models, we cannot increase the size of vocabularies unlimitedly. A large vocabulary also leads to data sparsity problems, which hurts model learning \cite{VOLT}.

Identifier splitting can decompose complex identifiers into several sub-words. Karampatsis et al. \cite{BigCode2020Karampatsis} find that on a corpus of around $11.6$ million unique tokens, the size of this corpus decrease dramatically by around $90\%$ after splitting identifiers. It implies that although the complex compounds are usually project-specific, the subunits that make up these compounds are highly repetitive across different projects. 
Instead of creating a vocabulary that contains many complex words, we propose to construct a vocabulary based on the corpus after identifier splitting (Split Corpora as in Figure~\ref{fig:pipeline}). 

We do not claim that using identifier splitting to process the corpus is always beneficial. Integrating identifier splitting in the vocabulary creation may lead to some negative impacts. The vocabulary created by the corpora after identifier splitting may need to use multiple sub-words to represent common compounds shared across different projects, e.g., popular API names. Splitting such common compounds may increase the length of the tokenized sequence and make it harder to relate the current prediction to the past context of inputs~\cite{BigCode2020Karampatsis}. Thus, we propose another strategy that merges the original and split corpora and builds a vocabulary from the combined corpora to tackle the issue (Merged Corpora as in Figure~\ref{fig:pipeline}). 

\vspace{0.1em}
\noindent{\bf Model Inputs Processing.} As shown in Figure~\ref{fig:pipeline}, the model input processing part takes input tokens and processes them into lists of sub-tokens for language model training.
To handle an input token that is not in the vocabulary, we use the BPE algorithm to decompose this input into a list of sub-tokens and then feed these sub-tokens into the model. As stated in Section \ref{sec:preliminary}, the vocabulary created by the BPE algorithm is an ordered list of sub-words. When BPE decomposes input tokens, it will follow the same order as recording in the vocabulary. We take the identifier {\tt getCategory} as an example. If we directly apply BPE (using a vocabulary of $10,000$ words) to this identifier, we get the following three sub-tokens: {\tt getC}, {\tt ateg} and {\tt ory}, which obviously break the original semantics of the identifier name. 
Although the word {\tt Category} is in vocabulary, its position ($9096th$) is almost close to the end. When BPE traverses the vocabulary, it will encounter and create the sub-token {\tt getC} ($1383th$) much earlier. If we first split this identifier into {\tt get} and {\tt Category}, and then apply BPE (using the same vocabulary) on the two words, we still get {\tt get} and {\tt Category}. Identifier splitting can utilize the semantic information conveyed with naming convention and prevent less meaningful sub-tokens (e.g., {\tt getC} and {\tt ateg}) from being created. This observation inspires us to apply identifier splitting before BPE.

Splitting model inputs may also lead to negative impacts. For instance, identifiers (e.g., types of exceptions or methods like {\tt toString}) can be shared across different projects, especially in object-oriented programming languages like Java. Such identifiers can frequently appear in the corpus and consequently are included in the vocabulary. They can be compactly represented only using one token, while identifier splitting will force them to be represented using multiple sub-tokens. As a result, we use a hybrid strategy to mitigate such negative impacts. More specifically, we first apply BPE to an identifier. If the tokenized result is identical to the original identifier, we directly feed it into the model. Otherwise, we feed the separated tokens to BPE after applying identifier splitting.

Considering the above, we combine the different operations in the vocabulary construction and the model inputs and propose the following three settings as shown in Figure~\ref{fig:pipeline} to explore the effectiveness of identifier splitting in the open-vocabulary LMs:
\begin{itemize}[leftmargin=*]
    \item {\bf Original:} using the BPE algorithm to create a vocabulary directly and then use the vocabulary to tokenize corpora as input. No identifier splitting is applied in this setting;
    \item {\bf Simple strategy:} splitting all identifiers in corpora first then use BPE to construct a vocabulary, and splitting all identifiers in model inputs;
    \item {\bf Hybrid strategy:} splitting identifiers and merging them with original corpora for BPE vocabulary construction, and splitting identifiers in model inputs only when BPE fails to tokenize them as the original forms.
\end{itemize}
\section{Experiments and Result Analysis} \label{sec:experiment}

\subsection{Implementation and Datasets}

To make the experiments under a computationally feasible scale, we select a lightweight, yet still effective Gated Recurrent Unit (GRU) model~\cite{GRU}. The model is also used in a recent work by Karampatsis et al.~\cite{BigCode2020Karampatsis}, which aims to analyze how the BPE algorithm can improve LMs of code. Also, inspired by~\cite{BigCode2020Karampatsis}, we limit the vocabulary size to 10k, set the input length and dimension of the GRU model as $200$ and $512$, and train all models using the stochastic gradient descent optimizer with a learning rate of $0.1$ and a mini-batch size of $32$.

We use the dataset released by Karampatsis et al.~\cite{BigCode2020Karampatsis} in our experiments, which consists of $177/141/73$ open-source projects in C language for training/validation/testing. Before training LMs, the corpora are processed in the same way as~\cite{BigCode2020Karampatsis}, removing strings of more than $15$ characters length, non-ASCII tokens and comments. We only use the training set to construct the vocabularies. 
The replication package are available via \url{https://github.com/soarsmu/CodeNLM.git}.

\subsection{Target Task and Evaluation Metrics} \label{subsec:metrics}

\textbf{Language Model.} 
We use the average per token cross entropy to evaluate the performance of our language models. The cross entropy is viewed as an intrinsic metric of LMs and employed in the previous work~\cite{BigCode2020Karampatsis}. By computing the logarithm mean of probability scores assigned by the LM over a sequence of source code, it estimates the average of bits required when using LMs to predict each token. A lower value of the cross entropy is favourable because it indicates LMs are easier to make correct predictions. Because the open-vocabulary LMs are based on sub-words units, the cross entropy is formalized as follows to compute the distribution over all sub-words $w_{i}^{1}, \dots, w_{i}^{m-1}$ instead of each token $t_i$:
\begin{equation}
    \small H(N) = -\frac{1}{N} \sum_{n=1}^N log \prod_{m=1}^M p(w_i^m|t_1,...,t_{n-1}, w_{i}^{1},..., w_{i}^{m-1})
\end{equation}
where $N$ is the number of tokens in the sequence, $M$ is the number of sub-words contained in $t_i$, and $p(w_i^m|t_1,...,t_{n-1}, w_{i}^{1},..., w_{i}^{m-1})$ is the probability of the sub-word $w_i^m$ given all the previous tokens and sub-words. 

\vspace{0.1em}
\textbf{Code Completion.} Automated code completion, an essential feature of modern integrated development environments (IDEs), aims to suggest a range of possible subsequent tokens within a toggle list. In open-vocabulary models, while a complete token could be a combination of multiple sub-words, this task can be formalized as maximizing the probability: $\arg\max p(w_1,\cdots,w_n|code\_before)$, in which $code\_before$ is the previous code snippet and $w_1,\cdots,w_n$ constitute the next complete token. The probability of the next complete token is the product of the probability of each sub-word. To obtain a complete token in open-vocabulary LMs, we use a customized beam search algorithm introduced by \cite{BigCode2020Karampatsis}, which can efficiently search through the sub-word expansion space and returns the top $k$ most possible complete candidates.

We evaluate the results with the widely-used {\em Mean Reciprocal Rank} (MRR) metric. MRR takes the rank of the correct answer as the primary grading criteria. For each token, if the correct answer ranks $n$th position among the top $k$ candidates, the score would be $\frac{1}{n}$. MRR is calculated by the following equation.
\begin{equation}
   \small MRR =  \frac{1}{|T|}\sum_{i = 1} ^{|k|} \frac{1}{rank_i}
\end{equation}
Statistics conducted by Hellendoorn et al. \cite{RealWorldCompletions} on real-world code completion scenarios observes that LMs for completion perform worse on identifiers than other types of tokens. Therefore, we also present the MRR and {\em recall at rank 10} (R@10) on predicting identifier tokens particularly (excluding keywords, punctuations, etc.).

\begin{table}[t!]
    \caption{Performance of the considered models under different strategies.}
    \label{tab:results}
    \centering
    \begin{scriptsize}
    \begin{tabular}{@{}ccccc@{}}
    \toprule
    \multirow{2}{*}{\textbf{Strategy}} &  \multicolumn{2}{c}{\textbf{All Tokens}}  & \multicolumn{2}{c}{\textbf{Identifiers}} \\ 
    \cmidrule(lr){2-3} \cmidrule(l){4-5}
    & Entropy & MRR &  R@10 & MRR \\ 
    \midrule
     Original & 4.46 & 64.61 & 37.55 & 21.83 \\
     Simple & 4.45(-0.22\%) & 64.31(-0.46\%)  & 36.26(-3.44\%) & 20.59(-5.68\%) \\
     Hybrid & \textbf{4.37(-2.02\%)} & \textbf{65.24(+0.98\%)}  & \textbf{38.93(+3.68\%)} & \textbf{23.19(+6.23\%)} \\ 
    \bottomrule
    \end{tabular}
    \end{scriptsize}
\end{table}

\subsection{Results}

We compare the performance of language models with different strategies. We denote the LMs with original settings, LMs with simple strategy and LMs with hybrid strategy as OriLMs, SSLMs and HSLMs, respectively.

Table~\ref{tab:results} shows the performance of different LMs. We find that SSLMs perform the worst among the three models. Although the entropy of SSLMs is slightly improved in comparison to OriLMs, SSLMs degrade the MRR on predicting all tokens by $0.46\%$, which reflects that the model's performance is not good as the OriLMs. For predicting identifier tokens, we observe a bigger performance gap between OriLMs and SSLMs. In terms of R@10 and MRR on predicting identifiers, OriLMs outperform SSLMs by $3.44\%$ and $5.68\%$, respectively. These results indicate that the simple strategy is not adequate for open-vocabulary LMs and even hurts the entropy and performance of LMs on the code completion task in most cases.

However, we observe that HSLMs outperform OriLMs by up to $0.98\%$ in terms of MRR on predicting all tokens. The models' entropy decreases $2.02\%$, showing that the hybrid strategy boosts open-vocabulary LMs. Compared with the results on all token prediction, HSLMs outperform OriLMs by a larger margin on identifier prediction: the results are boosted to $3.68\%$ and $6.23\%$ in terms of R@10 and MRR, respectively. The results demonstrate that combining identifier splitting with the BPE algorithm can improve the performance of open-vocabulary LMs, especially the improved performance on identifier prediction reveals that the LMs with hybrid strategy can synthesize identifiers from sub-words better.

In summary, by following the hybrid strategy, identifier splitting can boost the performance of open-vocabulary LMs of code. The performance of open-vocabulary LMs can be improved by $0.98\%$ and $6.23\%$ in terms of MRR on predicting both all tokens and identifiers. The entropy of LMs and R@10 on predicting identifiers can also be improved by $2.02\%$ and $3.68\%$.

\section{Related Work} \label{sec:related_work}

In the literature, prior studies have been interested in employing language models on source code to assist software development \cite{Raychev2014statistical, s2s}. However, the identifiers with complex names in source code make these models suffer from the {\em out-of-vocabulary} (OOV) problem. Increasing the size of a vocabulary has a limited effect on addressing the problem and makes models harder to scale \cite{VOLT}.
Recently, researchers have applied open vocabulary methods for code modeling. Karampatsis et al. \cite{BigCode2020Karampatsis} are the first to investigate whether open-vocabulary methods can improve the performance of this code completion tool. They trained a GRU-based language model with the BPE algorithm and showed that the model performance increases over close-vocabulary models across three datasets.

At the same time, numerous works about identifier splitting have been proposed before, and several studies have empirically compared these different techniques~\cite{Samurai, empirical2014split}. Some previous works also try to segment identifiers by naming conventions~\cite{Allamanis2015Suggesting, alon2018codeseq} in vocabulary construction, but no previous work utilizes advanced identifier splitting techniques and combines them with existing open-vocabulary algorithms.

\section{Conclusion and Future Work}
\label{sec:conclusion}

In this paper, we investigate the benefit of identifier splitting techniques on code modeling. We propose two strategies to combine identifier splitting with {\em Byte-Pair Encoding} (BPE) algorithm. We train open-vocabulary models with different strategies and compare the performance over the C language dataset. The evaluation results show that splitting identifiers improves the performance of open-vocabulary models under a hybrid strategy, which can improve LMs by $6.23\%$ in terms of MRR on predicting identifiers. The entropy of LMs and R@10 on predicting identifiers are also improved by up to $2.02\%$ and $3.68\%$. Our study confirms that the potential benefits of identifier splitting methods on open-vocabulary language models for C language.

In the future, we plan to validate our findings on more programming languages beyond C, e.g., Java and python. Also, we are interested in considering more models with different architectures, e.g., Transformer-based models, which have recently drawn researchers' attention. Besides, we plan to investigate whether the advanced LMs can boost more code modeling-based tasks, such as code clone detection~\cite{9516896}, bug localization~\cite{bl,incbl}, and code search~\cite{7081874}.

\section*{Acknowledgment}

This research / project is supported by the National Research Foundation, Singapore, under its Industry Alignment Fund – Pre-positioning (IAF-PP) Funding Initiative. Any opinions, findings and conclusions or recommendations expressed in this material are those of the authors and do not reflect the views of National Research Foundation, Singapore.

\bibliographystyle{IEEEtran}
\bibliography{ref}

\end{document}